\documentclass[11pt]{article}

\pdfoutput=1

\usepackage{times}
\usepackage{bm}
\usepackage{graphicx}
\usepackage{amsfonts}
\usepackage{amssymb}
\usepackage{amsmath}
\usepackage{graphicx}
\usepackage{natbib}
\usepackage{color}
\usepackage{graphicx}

\newtheorem{theorem}{Theorem}[section]
\newtheorem{proposition}{Proposition}[section]
\newtheorem{corollary}{Corollary}[section]
\newtheorem{lemma}{Lemma}[section]

\newcommand{\bproof}{\noindent\textsc{\bf Proof} \\}
\newcommand{\eproof}{$\Box$ \\}	

\newcommand{\vs}{\vspace{0.5cm}}
\renewcommand{\vss}{\vspace{0.25cm}}
\newcommand{\ds}{\displaystyle}

\usepackage{geometry}
\usepackage{setspace} 

\begin{document}

\title{A new semi-parametric family of probability distributions for survival analysis}
\date{}

\author{
{\bf Damien Bousquet}\footnote{\textsc{damien.bousquet@inserm.fr}} \\
Laboratoire de Biostatistiques, \'Epid\'emiologie, et de Sant\'e Publique \\
IURC, Universit\'e de Montpellier 1
\and
{\bf Jean-Pierre Daur\`es}\footnote{\textsc{jean-pierre.daures@inserm.fr}} \\
Laboratoire de Biostatistiques, \'Epid\'emiologie, et de Sant\'e Publique \\
IURC, Universit\'e de Montpellier 1
\and
{\bf Jean-Michel Marin}\footnote{place Eug\`ene Bataillon, CC051, 34095 Montpellier cedex 5, France, \textsc{jean-michel.marin@univ-montp2.fr} (corresponding author)} \\
Institut de Math\'ematiques et de Mod\'elisation de Montpellier \\
Universit\'e de Montpellier 2
}

\maketitle

\begin{abstract}

In the context of survival analysis, \cite{Marshall:Olkin:1997} introduced families of distributions
by adding a scalar parameter to a given survival function, parameterized or not.
In that paper, we generalize their approach. We show how it is possible to add more than a single parameter
to a given distribution. We then introduce very flexible families of distributions for which we calculate
some moments. Notably, we give some tractable expressions of these moments when the given baseline
distribution is Log-logistic. Finally, we demonstrate how to generate sample from these new families.

\vspace{0.5cm}  \noindent \textsc{Keywords}:  survival analysis; semi-parametric probability distributions;
Log-logistic and Weibull distributions.

\end{abstract}

\clearpage

\doublespacing

\section{Introduction}

By various methods, new parameters can be introduced to expand families of probability distributions.
This is an important issue in survival analysis \citep{Lawless:2003,Lee:Wang:2003,Marshall:Olkin:2007}.
For instance, although the Weibull distribution is often described as flexible, its hazard function
is restricted to being monotonically increasing or monotonically
decreasing, or constant. Typically, in survival analysis, the limitations of standard distributions led
naturally to interest in developing extended distributions by adding further parameters to a given distribution.

The approach of \cite{Marshall:Olkin:1997} works as follows. Let $S_0(x)$ denotes the survival function of a
random variable, the corresponding Marshall-Olkin extended distribution has survival function,
$$
\frac{a S_0(x)}{1-(1-a)S_0(x)}
$$
where $a>0$ is an added scalar parameter. These distributions have been used in many areas: 
\begin{itemize}
\item climatology: \cite{Biondi:Kozubowski:Panorska:Saito:2008};
\item hydrology: \cite{Jose:Naik:Ristic:2008};
\item insurance and finance: \cite{Garcia:Gomez-Deniz:Vazquez-Polo:2010}, \cite{Jayakumar:Mathew:2008} and \cite{Kozubowski:Panorska:2008};
\item medicine: \cite{Economou:Caroni:2007}, \cite{Ghitany:Al-Awadhi:Alkhalfan:2007}, \cite{Ghitany:Al-Hussaini:Al-Jarallah:2005}, \cite{Gomez-Deniz:2009},
\cite{Gupta:Peng:2009} and \cite{Jose:Naik:Ristic:2008};
\item engineering: \cite{Adamidis:Dimitrakopoulou:Loukas:2005}, \cite{Adamidis:Loukas:1998},
\cite{Gupta:Lvin:Peng:2010}, \cite{Prabhakar_Murthy:Bulmer:Eccleston:2004}, \cite{Silva:Barreto-Souza:Cordeiro:2010} and  \cite{Zhang:Xie:2007}...
\end{itemize}
For a review on the use of the extended Marshall and Olkin distributions, one can see \cite{Nadarajah:2008}.
There is also a lot of methodological works around the Marshall and Olkin extended distributions.
Independently of \cite{Marshall:Olkin:1997}, \cite{Adamidis:Loukas:1998} introduced the same type of extended distributions but only
in the baseline Weibull case. In \cite{Adamidis:Dimitrakopoulou:Loukas:2005}, it is shown that the Marshall and Olkin extended distributions
can be viewed as continuous mixtures. The baseline exponential case has been investigated by \cite{Silva:Barreto-Souza:Cordeiro:2010}
and \cite{Srinivasa-Rao:Ghitany:Kantam:2009}. For the more general baseline Weibull case, one can see
\cite{Ghitany:Al-Hussaini:Al-Jarallah:2005,Zhang:Xie:2007,Gupta:Lvin:Peng:2010}. \\
In \cite{Sankaran:Jayakumar:2008}, it is shown that the Marshall and Olkin extended distributions satisfy the property of proportional
odds functions. This property has been used notably by \cite{Economou:Caroni:2007}, \cite{Caroni:2008} and \cite{Gupta:Peng:2009} to introduce covariates. \\
Some authors have investigated the behavior of the Marshall and Olkin extended distributions when the baseline distributions are 
Lomax, \cite{Ghitany:Al-Awadhi:Alkhalfan:2007}, Burr, and Pareto, \cite{Jayakumar:Mathew:2008}, $q$-Weibull, \cite{Jose:Naik:Ristic:2008}
and normal \cite{Garcia:Gomez-Deniz:Vazquez-Polo:2010} distributions. Moreover, using the Marshall and Olkin approach,
\cite{Gomez-Deniz:2009} introduced a generalization of the discrete Geometric distribution. \\
In the baseline exponential case, \cite{Kozubowski:Panorska:2008} highlighted the link between the truncated logistic distribution
and the Marshall and Olkin extended one. \\
Finally, some authors have investigated the generalization to the multivariate case, one can see for instance:
\cite{Thomas:Jose:2004}, \cite{Sankaran:Jayakumar:2008}, \cite{Jose:Ristic:Joseph:2009}, \cite{Yeh:2009} and \cite{Yeh:2010}.
Random minima and maxima related to Marshall-Olkin distributions have received attention in the litterature, \cite{Arnold:1996}, \cite{Biondi:Kozubowski:Panorska:Saito:2008}, 
\cite{Ghitany:Al-Awadhi:Alkhalfan:2007}, \cite{Jose:Ristic:Joseph:2009}, \cite{Thomas:Jose:2004}, \cite{Yeh:2009} and \cite{Yeh:2010}.

\begin{figure}
\begin{center}
\includegraphics[width=0.3\textwidth]{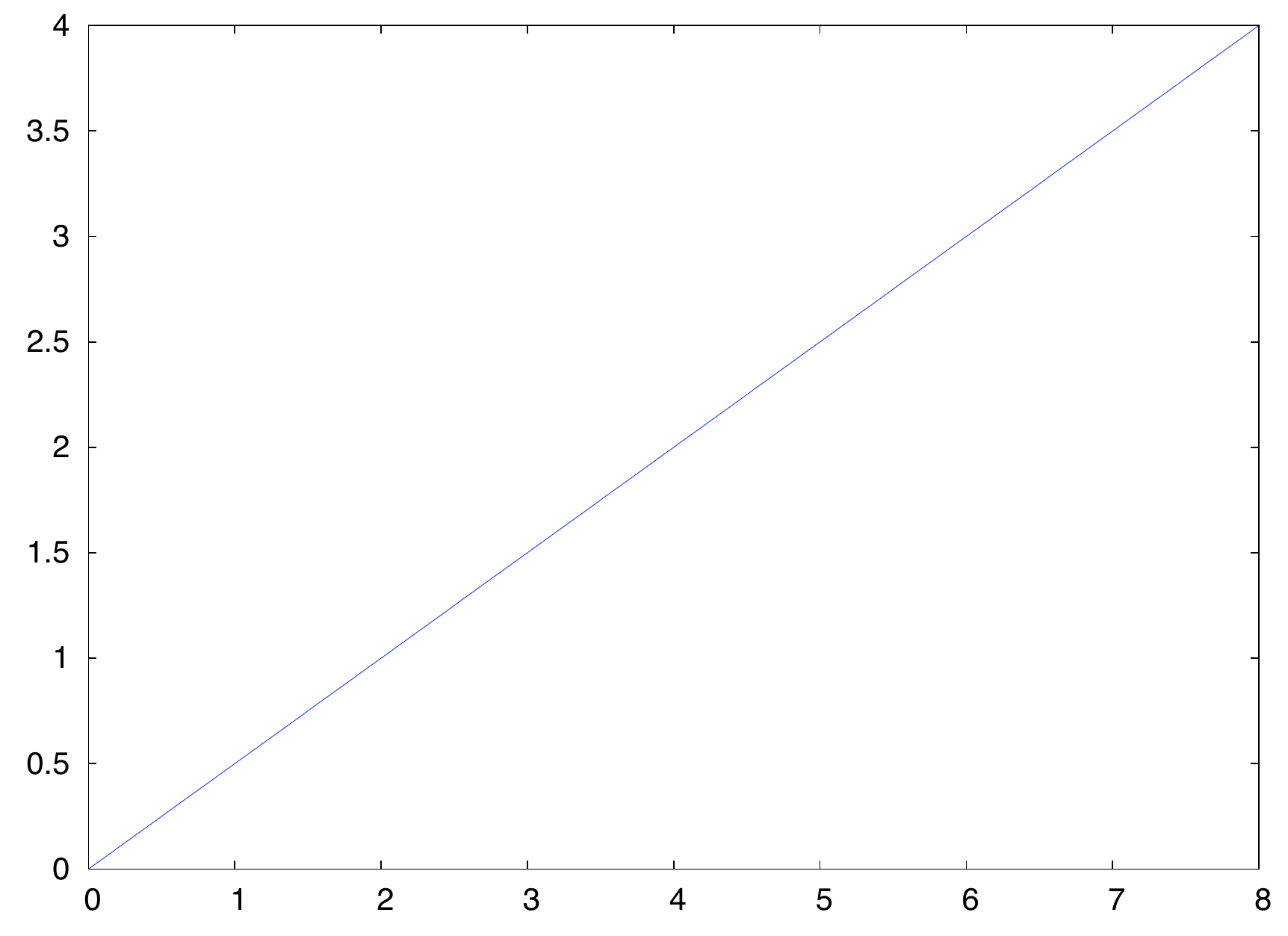}
\includegraphics[width=0.3\textwidth]{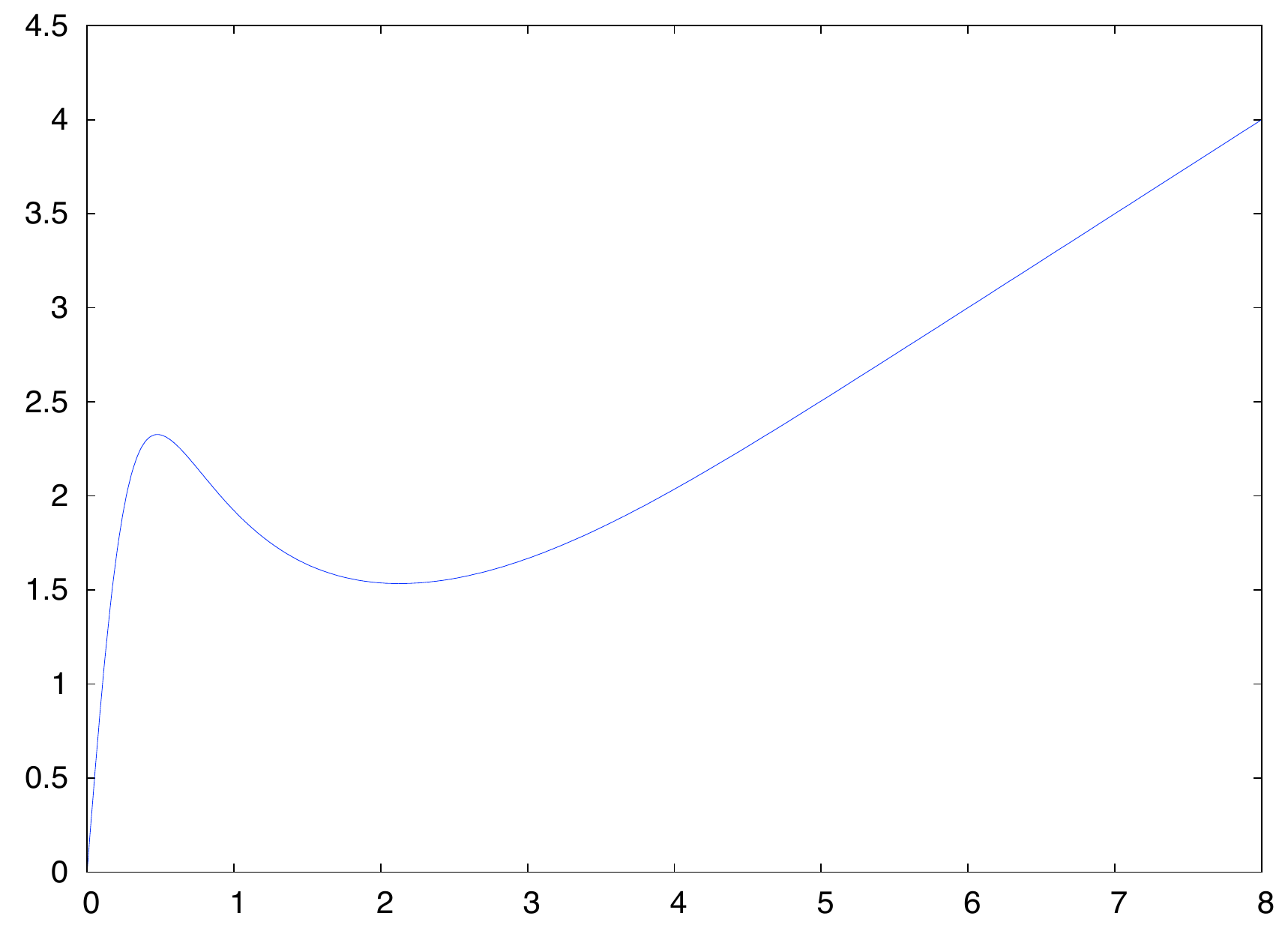}
\includegraphics[width=0.3\textwidth]{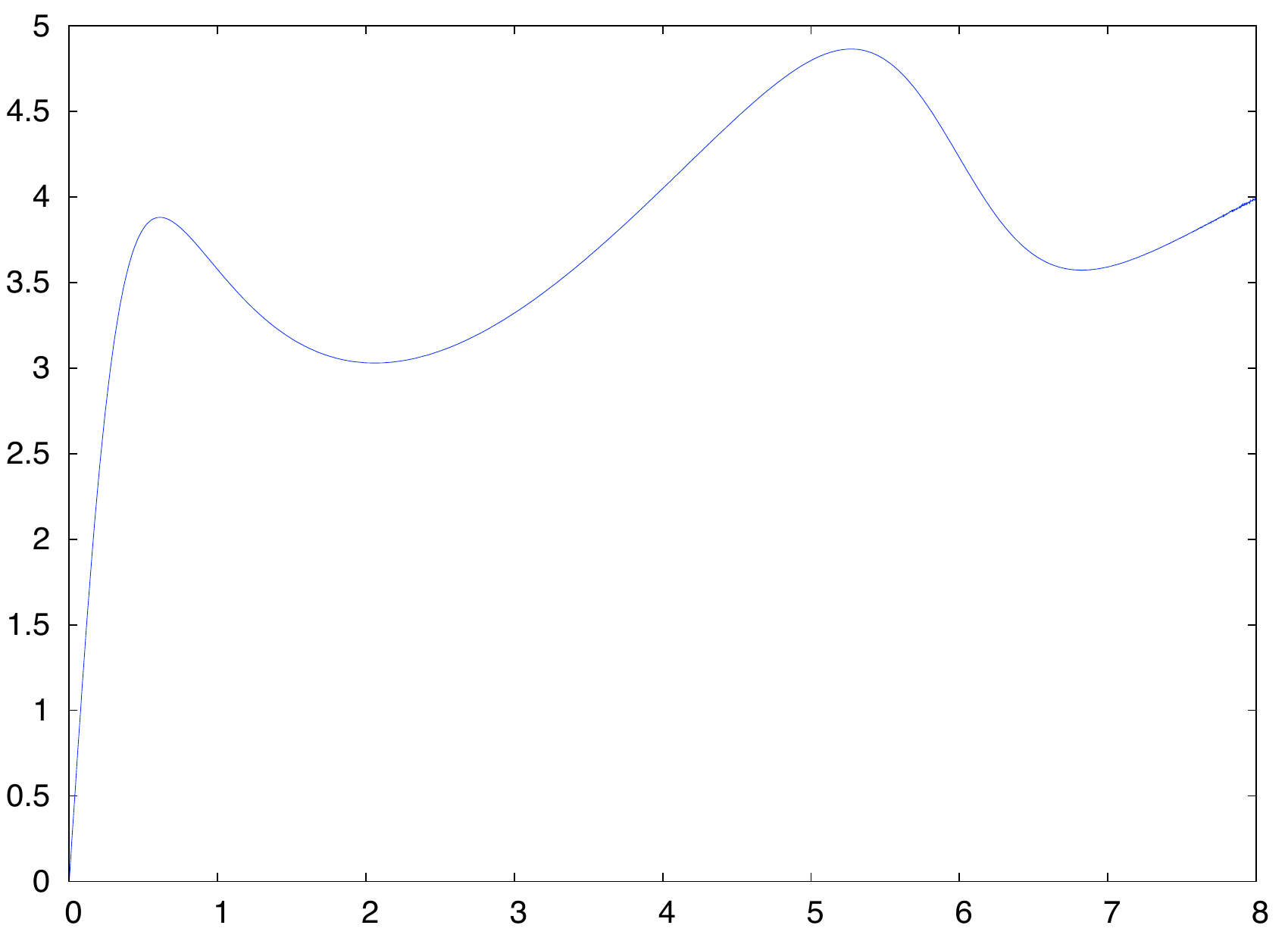}
\end{center}
\caption{Hazard rate curve of a standard Weibull distribution with parameter $(2,2)$ (left), of an extended Marshall and Olkin distribution
with added parameter $a=0.05$ (baseline Weibull $(2,2)$) (center) and of an extended distribution obtained using our approach:
two parameters ($a_1=10^{-6}$ and $a_2=0.15$) are added to a baseline Weibull $(2,2)$ (right).}
\label{figure:hazard}
\end{figure}

In that paper, we generalize the approach of Marshall and Olkin in order to obtain more flexible distributions.
Indeed, coming back to baseline Weibull one, it is easy to show that the hazard rate curve of the extended Marshall and Olkin distribution
has at most two waves. As shown in Figure \ref{figure:hazard}, using our approach, we can obtain more rich hazard rate curve.
Our proposal is introduced in Section \ref{section:nf}. In Section \ref{section:properties}, we investigate some properties
of these extended distributions for various baseline cases.

\section{The new semi-parametric family}
\label{section:nf}

In~\cite{Marshall:Olkin:1997}, it is introduced a new semi-parametric family of probability distributions.
One real parameter $a>0$ is added to a given univariate probability distribution $F_0:\mathbb{R} \rightarrow [0,1]$
by using an increasing function denoted by $g_a$. Indeed, the Marshal and Olkin
extended probability distribution is given by
$$
g_a\left(F_0(x)\right)=\frac{F_0(x)}{a+(1-a)F_0(x)}\,.
$$
Here, we generalize this support function with an arbitrary number of external parameters.

Let $S_0=1-F_0$ and $q\in\mathbb{N}^*$.
Clearly, for any $(a_1,\ldots,a_q) \in {]0,+\infty[}^q$ and any $u\in[0,1]$, we have,
$$
\left\{\sum_{i=1}^q a_i-\left(\sum_{i=1}^q a_i-q\right)u\right\} \neq 0\,.
$$
We define the function $g_{a_1,\ldots,a_q}:[0,1]\rightarrow \mathbb{R}$ by
$$
g_{a_1,\ldots,a_q}(u)=q^q \frac{\ds u \prod_{i=2}^q (a_i+u-a_i u)}{\ds {\left\{\sum_{i=1}^q a_i-\left(\sum_{i=1}^q a_i-q\right)u\right\}}^q}\,.
$$
As a generalization of the Marshal and Olkin proposal, we define the 
Let us now define the functions $F_{a_1,\ldots,a_q}$ and $S_{a_1,\ldots,a_q}$ such that :
$$
\begin{array}{cccl}
F_{a_1,\ldots,a_q}: & \mathbb{R} & \rightarrow & \mathbb{R} \\
                    & x          & \mapsto     & g_{a_1,\ldots,a_q}\left\{F_0(x)\right\}=q^q \frac{F_0(x) \prod_{i=2}^q 
\left\{a_i+F_0(x)-a_i F_0(x)\right\}}{{\left\{\sum_{i=1}^q a_i-(\sum_{i=1}^q a_i-q) F_0(x)\right\}}^q}\,\,;
\end{array}
$$

\begin{theorem}
\label{theorem:mainres}
For any integer $q\in\mathbb{N}^*$ and any $(a_1,\ldots,a_q) \in {]0,+\infty[}^q$, the application $F_{a_1,\ldots,a_q}$ from $\mathbb{R}$ to $\mathbb{R}$ takes its values in $[0,1]$, 
and the function $F_{a_1,\ldots,a_q}$ from $\mathbb{R}$ to $[0,1]$ defines a probability distribution.
\end{theorem}

\bproof
The proof is given in Appendix.
\eproof

If we replace in Theorem~\ref{theorem:mainres}, the cumulative probability function $F_0$ by the survival function $S_0$, we still obtain a survival function.
This defines a new  probability distribution. In that work, we only consider the initial construction.

\newpage

For any $a>0$, the function $F_{a,\ldots,a}$ from $\mathbb{R}$ to $[0,1]$ is a Marshall-Olkin extended probability distribution.
That is, for any real number $x$, we have,
$$
F_{a,\ldots,a}(x)=\frac{F_0(x)}{a+(1-a)F_0(x)}\,,
$$
or, in another way,
$$
S_{a,\ldots,a}(x)=\frac{a S_0(x)}{1-(1-a)S_0(x)}\,.
$$

\vs Let us now introduce the following lemma.

\begin{lemma}
\label{lemma:support}
We suppose that there exists $t \in \mathbb{R}$ such that $0<F_0(t)<1$. \\
In that case, there exists a unique interval $J$ of $\mathbb{R}$ such that, for any $x \in (\mathbb{R}-J) \cap ]-\infty,t]$ we have
$F_0(x)=0$, for any $x \in (\mathbb{R}-J) \cap [t,+\infty[$ we have $F_0(x)=1$, and for any $x$ in $J$ we have
$0<F_0(x)<1$.
\end{lemma}

The proof of this lemma is trivial. The interval $J$ does not correspond to the usual definition of the support
of a probability distribution (the smallest closed set whose complement has probability zero).
We use that specific definition for technical reasons.

\vs Let us suppose that $F_0$ is continuous on $\mathbb{R}$ and $C^1$ on $J$, the interval $J$ is defined by Lemma \ref{lemma:support}. \\
Let $f_0$ be the probability density function of $F_0$. We can take $f_0$ as,
$$
\begin{array}{ccccl}
f_0: & \mathbb{R} & \rightarrow & \mathbb{R}_+ \\
     & x          & \mapsto     & F_0'(x)     & \mathrm{if} \; x \in J \\
     &            &             & 0           & \mathrm{otherwise}
\end{array}
$$
Here, it is very easy to verify that for any $q\in\mathbb{N}^*$ and any $(a_1,\ldots,a_q) \in {]0,+\infty[}^q$,
the function $F_{a_1,\ldots,a_q}$ from $\mathbb{R}$ into $[0,1]$ is continuous on $\mathbb{R}$ and $C^1$ on $J$. \\
Moreover, the probability density function $f_{a_1,\ldots,a_q}$ of $F_{a_1,\ldots,a_q}$ can be taken as,  
$$
\begin{array}{ccccl}
f_{a_1,\ldots,a_q}: & \mathbb{R} & \rightarrow & \mathbb{R}_+ \\
                    & x          & \mapsto     & F_{a_1,\ldots,a_q}'(x)=g_{a_1,\ldots,a_q}'\left\{F_0(x)\right\} f_0(x) & \mathrm{if} \; x \in J \\
                    &            &             & 0                                                                      & \mathrm{otherwise}
\end{array}
$$ 
For instance if $q=2$, we get for any $x \in J$,
$$
f_{a_1,a_2}(x)=4 \frac{a_1a_2F_0(x)+a_1^2F_0(x)-a_1a_2-a_1^2-2a_2F_0(x)}{{\left\{a_1F_0(x)+a_2F_0(x)-a_1-a_2-2F_0(x)\right\}}^3} f_0(x)\,.
$$

Let us now introduce two Lemmas.

\begin{lemma}
\label{lemma:dev1}
Let $q\in\mathbb{N}^*$ and $a_1, \ldots,a_q$ be strictly positive real numbers such that \\
$a_1+\cdots+a_q > \frac{q}{2}$. For any $u\in[-1,1]$,
$$
\sum_{i=1}^q a_i-\left(\sum_{i=1}^q a_i-q\right)u \neq 0\,,
$$
and
$$
g_{a_1,\ldots,a_q}(u)=\sum_{m=1}^{\infty} c_{q,a_1,\ldots,a_q,m} u^m
$$
where, for any $m\in\mathbb{N}^*$,
$$
c_{q,a_1,\ldots,a_q,m}=
$$
$$
\frac{q^q}{(q-1)!} \frac{\prod_{i=2}^q a_i}{{(\sum_{i=1}^{q} a_i)}^q} \left\{ \sum_{j+k=m,1 \leq j \leq q} \sigma_{j-1} \; (k+q-1) \cdots (k+1) 
\left(\frac{\sum_{i=1}^{q} a_i-q}{\sum_{i=1}^{q} a_i}\right)^k \right\}
$$
with
$$
\sigma_i=\left\{\begin{array}{ll}
1 & \mathrm{if}\quad i=0 \\
\sum_{2 \leq j_1 < \ldots <j_i \leq q} \frac{1-a_{j_1}}{a_{j_1}} \times \cdots \times \frac{1-a_{j_i}}{a_{j_i}} & \mathrm{if}\quad 1\leq i \leq q-1 \\
0 & \mathrm{if} \quad i\geq q
\end{array}\right\}\,.
$$
Note that we have
$$
\sum_{m=1}^{\infty} c_{q,a_1,\ldots,a_q,m}=1\,.
$$
\end{lemma}

\bproof
The proof is given in Appendix.
\eproof

We trivially deduce the following corollary used in the proof of Theorem \ref{theorem:expectation}.

\begin{corollary}
\label{corollary:signe_coefficients}
Let $q\in\mathbb{N}^*$ and $a_1, \ldots,a_q$ be strictly positive real numbers
such that $a_1+\cdots+a_q \geq q$ and $a_i \leq 1$ for $2 \leq i \leq q$. Then, for any $m\in\mathbb{N}^*$,
$$
c_{q,a_1,\ldots,a_q,m} \geq 0\,.
$$
\end{corollary}

\begin{lemma}
\label{lemma:dev2}
Let $q\in\mathbb{N}^*$ and $a_1, \ldots,a_q$ be strictly positive real numbers such that \\
$a_1+\cdots+a_q < 2q$. For any $u\in[0,2]$,
$$
\sum_{i=1}^q a_i-\left(\sum_{i=1}^q a_i-q\right)u \neq 0\,,
$$
and
$$
g_{a_1,\ldots,a_q}(u)=1+\sum_{m=1}^{\infty} d_{q,a_1,\ldots,a_q,m} {(u-1)}^m
$$
where, for $m\in\mathbb{N}^*$,
\begin{eqnarray*}
d_{q,a_1,\ldots,a_q,m}= & \frac{1}{(q-1)!} \left\{ (m+q-1) \cdots (m+1) {(\frac{\sum_{i=1}^{q} a_i-q}{q})}^m \right.\\
& \left.+\sum_{k+j=m,1 \leq j \leq q} (k+q-1) \cdots (k+1) {(\frac{\sum_{i=1}^{q} a_i-q}{q})}^k (\sigma_j+\sigma_{j-1}) \right\}
\end{eqnarray*}
with
$$
\sigma_i=\left\{\begin{array}{ll}
1 & \mathrm{if}\quad i=0 \\
\sum_{2 \leq j_1 < \ldots <j_i \leq q} (1-a_{j_1}) \times \cdots \times (1-a_{j_i}) & \mathrm{if}\quad 1\leq i \leq q-1 \\
0 & \mathrm{if} \quad i\geq q
\end{array}\right\}\,.
$$
\end{lemma}

\bproof
The proof is given in Appendix.
\eproof

We now study the expectation of our family of distributions when the baseline probability distribution $F_0$
is continuous on $\mathbb{R}$ and $C^1$ on $J$.
 
\begin{theorem}
\label{theorem:expectation}
Let $(\Omega,\mathcal{A},P)$ be a probability space, $q\in\mathbb{N}^*$, $(a_1,\ldots,a_q) \in {]0,+\infty[}^q$, \\ 
$X_0:\Omega \rightarrow \mathbb{R}$ and $X_{a_1,\ldots,a_q}:\Omega \rightarrow \mathbb{R}$ be random variables
such that $X_0\sim F_0$ and $X_{a_1,\ldots,a_q}\sim F_{a_1,\ldots,a_q}$ and
$w: \mathbb{R} \rightarrow \mathbb{R}$ be a borelian function.

If $F_0$ is continuous on $\mathbb{R}$ and $C^1$ on $J$ and if $E(|w \circ X_0|)=\int_{\mathbb{R}} |w(x)| f_0(x) dx \in \mathbb{R}_+$, then
$$
E(|w \circ X_{a_1,\ldots,a_q}|) \in \mathbb{R}_+\,.
$$
Moreover, if $a_1,\ldots,a_q$ be strictly positive real numbers such that $a_1+\cdots+a_q \geq q$ and $a_i\leq 1$ for any $2 \leq i \leq q$, then
$$
E(|w \circ X_{a_1,\ldots,a_q}|) \leq a_1 E(|w \circ X_0|)\,.
$$
\end{theorem}

\bproof
The proof is given in Appendix.
\eproof

\newpage

As a direct consequence,  if $a_1,\ldots,a_q$ be strictly positive real numbers such that $a_1+\cdots+a_q \geq q$
and $a_i\leq 1$ for any $2 \leq i \leq q$ and, $\int_{\mathbb{R}} {|x|}^r f_0(x) dx \in \mathbb{R}_+$ for a real number $r$, then 
$$
E({|X_{a_1,\ldots,a_q}|}^r) \leq a_1 E({|X_0|}^r)\,.
$$

The following proposition will be used to derive explicit formulae on some expectations when
the baseline probability distribution is a classical Log-logistic.

\begin{proposition}
\label{proposition:convolution}
We suppose that $F_0$ is bijective from $J$ to $]0,1[$. Let $(\Omega,\mathcal{A},P)$ be a probability space,
$(a_1,a_2) \in {]0,+\infty[}^2$ such that $a_1 \neq a_2$,
$X_{a_1,a_2}:\Omega \rightarrow \mathbb{R}$ and $V_{a_1,a_2}:\Omega \rightarrow \mathbb{R}$ be independent random variables
such that $X_{a_1,a_2}\sim F_{a_1,a_2}$ and  $V_{a_1,a_2}\sim \mathrm{Exp}\left(\frac{a_1+a_2}{|a_1-a_2|}\right)$.

Define for any $\omega \in \Omega$ such that $X_{a_1,a_2}(\omega) \in J$,
$$
Y_{a_1,a_2}(\omega)=\log\left[\frac{1}{F_0\{X_{a_1,a_2}(\omega)\}}-1\right]\,,
$$
and for any $\omega \in \Omega$ such that $X_{a_1,a_2}(\omega) \in \mathbb{R}-J$,
$$
Y_{a_1,a_2}(\omega)=0\,.
$$
Then the random variable,
$$
L_{a_1,a_2}=\left\{\begin{array}{ll}
 Y_{a_1,a_2}+V_{a_1,a_2} -\log \frac{2}{a_1+a_2}  & \mathrm{if}\quad a_1>a_2 \\
 -Y_{a_1,a_2}+V_{a_1,a_2} -\log \frac{a_1+a_2}{2} & \mathrm{if}\quad a_1<a_2
\end{array} \right.
$$
is distributed according to a logistic distribution.
\end{proposition}

\newpage

\bproof
Let us denote by $\alpha_{a_1,a_2}$ the cumulative probability function of $V_{a_1,a_2}$, by $r_{a_1,a_2}$ the one of $Y_{a_1,a_2}$,
by $t_{a_1,a_2}$ the one of $-Y_{a_1,a_2}$, and by $u_{a_1,a_2}$ the one of $L_{a_1,a_2}$. 

For any $u \in\mathbb{R}$, we have
$$
r_{a_1,a_2}(u)=1-g_{a_1,a_2}\left\{\frac{1}{1+\exp(u)}\right\}
$$
and
$$
t_{a_1,a_2}(u)=g_{a_1,a_2}\left\{\frac{\exp(u)}{1+\exp(u)}\right\}\,.
$$

\vss We first suppose that $a_1>a_2$ and for any $(u,t) \in {\mathbb{R}}^2$, we define
$$
\theta_{a_1,a_2,u}(t)=-\frac{2(a_1+a_2) \exp(u+\frac{2a_2}{a_2-a_1}t)}{{\left\{2\exp(t)+(a_1+a_2)\exp(u)\right\}}^2}\,.
$$
For any $(u,t) \in \mathbb{R} \times [0,+\infty[$, we clearly have
$$
\theta_{a_1,a_2,u}'(t)=r_{a_1,a_2}'(u-t)\alpha_{a_1,a_2}'(t)
$$
and
$$
\int_{\mathbb{R}_+} r_{a_1,a_2}'(u-t)\alpha_{a_1,a_2}'(t) dt = \frac{2(a_1+a_2)\exp(u)}{{\left\{2+(a_1+a_2)\exp(u)\right\}}^2}\,.
$$
Therefore, for any $v\in\mathbb{R}$,
$$
\int_{-\infty}^v \int_{\mathbb{R}_+} r_{a_1,a_2}'(u-t)\alpha_{a_1,a_2}'(t) dt du=\frac{(a_1+a_2)\exp(v)}{2+(a_1+a_2)\exp(v)}=u_{a_1,a_2}\left(v-\log \frac{2}{a_1+a_2}\right)
$$
and
$$
u_{a_1,a_2}(v)=\frac{\exp(v)}{1+\exp(v)}\,.
$$
We now suppose that $a_2>a_1$ and for any $(u,t) \in {\mathbb{R}}^2$ and we define 
$$
\theta_{a_1,a_2,u}(t)=-\frac{2(a_1+a_2)\exp(u-\frac{2a_1}{a_2-a_1}t)}{{\left\{2\exp(u)+(a_1+a_2)\exp(t)\right\}}^2}\,.
$$
For any $(u,t) \in \mathbb{R} \times [0,+\infty[$, we clearly have
$$
\theta_{a_1,a_2,u}'(t)=t_{a_1,a_2}'(u-t)\alpha_{a_1,a_2}'(t)
$$
and
$$
\int_{\mathbb{R}_+} t_{a_1,a_2}'(u-t)\alpha_{a_1,a_2}'(t) dt = \frac{2(a_1+a_2) \exp(u)}{{\left\{2\exp(u)+a_1+a_2\right\}}^2}\,.
$$
Therefore, for any $v\in\mathbb{R}$,
$$
\int_{-\infty}^v \int_{\mathbb{R}_+} t_{a_1,a_2}'(u-t)\alpha_{a_1,a_2}'(t) dt du=\frac{2\exp(v)}{2\exp(v)+a_1+a_2}=u_{a_1,a_2}\left(v-\log \frac{a_1+a_2}{2}\right)
$$
and
$$
u_{a_1,a_2}(v)=\frac{\exp(v)}{1+\exp(v)}\,.
$$
\eproof

In~\cite{Marshall:Olkin:1997}, it is shown a specific property of the introduced semi-parametric family:
the proposed distributions are geometric extreme stable. That is, they are both minimum and maximum stable, when the random indexation variable of the sample size 
is distributed according to a geometric distribution. We now consider maximum stability for our semi-parametric family.

\newpage

\begin{proposition}
Let $q\in\mathbb{N}^*$ and $a_1,\ldots,a_q$ be strictly positive real numbers such that $a_1+\cdots+a_q \geq q$ and $a_i\leq 1$ for any $2 \leq i \leq q$. 
Let $(\Omega,\mathcal{A},P)$ be a probability space and $N_{a_1,\ldots,a_q}:\Omega \rightarrow \mathbb{N}^*$ be a random variable such that, 
for any $m\in\mathbb{N}^*$,
$$
\mathrm{pr}(N_{a_1,\ldots,a_q}=m)=c_{q,a_1,\ldots,a_q,m}\,.
$$
\begin{itemize}
\item If, for $i \in \mathbb{N}^*$, $X_i: \Omega \rightarrow \mathbb{R}$ be mutually independent random variables, independent from $N_{a_1,\ldots,a_q}$,
such that $X_i\sim F_0$, then 
$$
V_{a_1,\ldots,a_q}=\max(X_1, \ldots,X_{N_{a_1,\ldots,a_q}})\sim F_{a_1,\ldots,a_q}\,.
$$
\item If, for $b>0$ and $i \in \mathbb{N}^*$, $X_i: \Omega \rightarrow \mathbb{R}$ be mutually independent random variables,
independent from $N_{a_1,\ldots,a_q}$ such that $X_i\sim F_{b,\ldots,b}$, then
$$
V_{a_1,\ldots,a_q}=\max(X_1, \ldots,X_{N_{a_1,\ldots,a_q}})\sim F_{b a_1,\ldots,b a_q}\,.
$$
\end{itemize}
\end{proposition}

\bproof
Part one: we have for any real number $x$,
$$
\mathrm{pr}(V_{a_1,\ldots,a_q} \leq x) = \sum_{m=1}^{\infty} \mathrm{pr}(N_{a_1,\ldots,a_q}=m) {F_0(x)}^m \,,
$$
$$
\mathrm{pr}(V_{a_1,\ldots,a_q} \leq x)=g_{a_1,\ldots,a_q}\left\{F_0(x)\right\} \,,
$$
$$
\mathrm{pr}(V_{a_1,\ldots,a_q} \leq x)=F_{a_1,\ldots,a_q}(x) \,.
$$

\vs Part two: for any $(u,a_1,\ldots,a_q,b) \in [0,1] \times {]0,+\infty[}^{q+1}$, we have,
$$
g_{a_1,\ldots,a_q}\left\{g_{b,\ldots,b}(u)\right\}=g_{b a_1,\ldots,b a_q}(u)\,.
$$
\eproof

More generally for $q \geq 2$, the new probability distributions are not maximum stable.

\section{Some properties of the extended distributions}
\label{section:properties}

\subsection{$F_0$ is Weibull}

We first suppose that $F_0(x)=\left(1-\exp(-x)\right)\mathbb{I}_{x\geq 0}$ (exponential distribution with expectation
equal to 1).
\begin{lemma}
\label{lemma:exp}
Let $F_0(x)=\left(1-\exp(-x)\right)\mathbb{I}_{x\geq 0}$ and $f_0$ is the corresponding probability density function.
For any $r>0$ and any $m\in\mathbb{N}^*$, we have
$$
\int_{\mathbb{R}} x^r {F_0(x)}^{m-1} f_0(x) dx=r \Gamma(r) \sum_{j=0}^{m-1} {m-1 \choose j} {(-1)}^j {(j+1)}^{-r-1}\,.
$$
\end{lemma}

\newpage

\bproof
Let $r>0$, $p\in\mathbb{N}$ and $z \geq 0$,
$$
{F_0(z)}^p=\sum_{j=0}^p {p \choose j} {(-1)}^j \exp(-jz)
$$
and
$$
z^r {F_0(z)}^p f_0(z)=\sum_{j=0}^p {p \choose j} {(-1)}^j z^r \exp\left\{-(j+1)z\right\}\,.
$$
Then,
$$
\int_{\mathbb{R}_+} x^r {F_0(x)}^p f_0(x) dx=\sum_{j=0}^p {p \choose j} {(-1)}^j \int_{\mathbb{R}_+} x^r \exp\left\{-(j+1)x\right\} dx
$$
and
$$
\int_{\mathbb{R}_+} x^r {F_0(x)}^p f_0(x) dx=\sum_{j=0}^p {p \choose j} {(-1)}^j \frac{r\Gamma(r)}{{(j+1)}^{r+1}} =r \Gamma(r) \sum_{j=0}^p {p \choose j} {(-1)}^j {(j+1)}^{-r-1}\,.
$$
\eproof
\begin{proposition}
\label{proposition:moments_exponentielle}
Let $F_0(x)=\left(1-\exp(-x)\right)\mathbb{I}_{x\geq 0}$ and $X_{a_1,\ldots,a_q}\sim F_{a_1,\ldots,a_q}$.

\vspace{0.3cm} If  $a_1,\ldots,a_q$ be strictly positive real numbers such that $a_1+\cdots+a_q \geq q$ and $a_i\leq 1$ for $2 \leq i \leq q$,
$\displaystyle E({X_{a_1,\ldots,a_q}}^r)=r \Gamma(r) \sum_{m=1}^{\infty} m c_{q,a_1,\ldots,a_q,m} \sum_{j=0}^{m-1} {m-1 \choose j} {(-1)}^j {(j+1)}^{-r-1}$ for any $r>0$.
 
\vspace{0.3cm} If $a_1,\ldots,a_q$ be strictly positive real numbers such that $q\leq a_1+\cdots+a_q < 2q$ and $a_i\leq 1$ for $2 \leq i \leq q$,
$\displaystyle E({X_{a_1,\ldots,a_q}}^r)=r \Gamma(r) \sum_{m=1}^{\infty} {(-1)}^{m-1} d_{q,a_1,\ldots,a_q,m} m^{-r}$ for any $r>0$.
\end{proposition}

\bproof
The proof of this result is almost trivial using Lemmas \ref{lemma:dev1}, \ref{lemma:dev2} and \ref{lemma:exp}.
\eproof


\vs We now consider the case of the classical Weibull distribution with two parameters, that is
$$
F_0(x)=F_0^{b_1,b_2}(x)=\left(1-\exp\left\{-\left(\frac{x}{b_1}\right)^{b_2}\right\}\right)\mathbb{I}_{x\geq 0}
$$
with $(b_1,b_2) \in ]0,+\infty[^2$. Let $X_{a_1,\ldots,a_q}^{b_1,b_2}\sim F_{a_1,\ldots,a_q}^{b_1,b_2}$ ($F_{a_1,\ldots,a_q}^{b_1,b_2}$
be the corresponding parameter augmented distribution).
We can easily verify that $X_{a_1,\ldots,a_q}^{b_1,b_2}=b_1 {X_{a_1,\ldots,a_q}^{1,1}}^{\frac{1}{b_2}}$ in distribution and, then for any $r>0$,
$$
E\left({X_{a_1,\ldots,a_q}^{b_1,b_2}}^r\right)={b_1}^r E\left({X_{a_1,\ldots,a_q}^{1,1}}^{\frac{r}{b_2}}\right)\,.
$$

\vs Finally, we consider the case of the generalized Weibull distribution with three parameters, that is 
$$
F_0^{b_1,b_2,b_3}(x)=\left(1-\exp\left[1-{\left\{1+{(\frac{x}{b_1})}^{b_2}\right\}}^{\frac{1}{b_3}}\right]\right)\mathbb{I}_{x\geq 0}
$$
with $(b_1,b_2,b_3) \in ]0,+\infty[^3$. Let $X_{a_1,\ldots,a_q}^{b_1,b_2,b_3}\sim F_{a_1,\ldots,a_q}^{b_1,b_2,b_3}$.
We can easily verify that \\ $X_{a_1,\ldots,a_q}^{b_1,b_2,b_3}=b_1 {\left\{{(1+X_{a_1,\ldots,a_q}^{1,1,1})}^{b_3}-1\right\}}^{\frac{1}{b_2}}$
in distribution and then, if we suppose in addition that $m_2=\frac{1}{b_2}\in\mathbb{N}-\{0\}$ and $b_3\in\mathbb{N}^*$,
it follows that for any integer $m\in\mathbb{N}^*$,
$$
E({X_{a_1,\ldots,a_q}^{b_1,b_2,b_3}}^m)=b_1^m \left\{\sum_{k=0}^{m m_2} \sum_{j=0}^{b_3 k} {m m_2 \choose k} {b_3k \choose j} {(-1)}^{m m_2-k} E({X_{a_1,\ldots,a_q}^{1,1,1}}^{kb_3j})\right\}\,.
$$

\newpage

\subsection{$F_0$ is Log-logistic}

We now consider the case of the standard Log-logistic distribution, that is
$$
F_0(x)=\left(\frac{x}{1+x}\right)\mathbb{I}_{x\geq 0}\,.
$$

\begin{proposition}
\label{proposition:moments_log_logistique}
Let $F_0(x)=\left(\frac{x}{1+x}\right)\mathbb{I}_{x\geq 0}$ and $X_{a_1,\ldots,a_q}\sim F_{a_1,\ldots,a_q}$.

\vspace{0.3cm} If $a_1,\ldots,a_q$ be strictly positive real numbers such that $a_1+\cdots+a_q \geq q$ and $a_i\leq 1$ for $2 \leq i \leq q$, then for any $r$ such that $|r|<1$,
$$
E({X_{a_1,\ldots,a_q}}^r)=\sum_{m=1}^{\infty} m c_{q,a_1,\ldots,a_q,m} \mathrm{Beta}(1-r,m+r)\,.
$$

\vspace{0.3cm} If $a_1,\ldots,a_q$ be strictly positive real numbers such that $q\leq a_1+\cdots+a_q < 2q$ and $a_i\leq 1$ for $2 \leq i \leq q$, then for any $r$ such that $|r|<1$,
$$
E({X_{a_1,\ldots,a_q}}^r)=\sum_{m=1}^{\infty} {(-1)}^{m-1} m d_{q,a_1,\ldots,a_q,m} \mathrm{Beta}(m-r,1+r)\,.
$$
\end{proposition}

\bproof
The proof of this result is almost trivial using Lemmas \ref{lemma:dev1} and \ref{lemma:dev2}.
\eproof
We now consider the case of the classical Log-logistic distribution with two parameters, that is
$$
F_0^{b_1,b_2}(x)=\left(\frac{x^{b_2}}{{b_1}^{b_2}+x^{b_2}}\right)\mathbb{I}_{x\geq 0}
$$
with $(b_1,b_2) \in ]0,+\infty[^2$.

Let $X_{a_1,\ldots,a_q}^{b_1,b_2}\sim F_{a_1,\ldots,a_q}^{b_1,b_2}$.
As in the Weibull case, we can easily to verify that $X_{a_1,\ldots,a_q}^{b_1,b_2}=b_1 {X_{a_1,\ldots,a_q}^{1,1}}^{\frac{1}{b_2}}$ in distribution,
and then for any $r$ such that $|r|<b_2$,
$$
E({X_{a_1,\ldots,a_q}^{b_1,b_2}}^r)={b_1}^r E({X_{a_1,\ldots,a_q}^{1,1}}^{\frac{r}{b_2}})\,.
$$

\begin{proposition}
Let $F_0\left(\frac{x}{1+x}\right)\mathbb{I}_{x\geq 0}$ and $X_{a_1,a_2}\sim F_{a_1,a_2}$. For any $r$ such that $|r|<1$, 
$$
E({X_{a_1,a_2}}^r)=\left(\frac{a_1+a_2}{2}\right)^r \frac{r\pi}{\sin(r\pi)} \left(\frac{a_1-a_2}{a_1+a_2}r+1\right)\,.
$$
\end{proposition}

\bproof
When the baseline distribution is a classical Log-logistic with two parameters, the extended Marshall-Olkin distribution with
one external parameter gives again a classical Log-logistic distribution. Therefore, the case $a_1=a_2$ is trivial.

\vss Let us now consider the case $a_1>a_2$
\\Using Proposition \ref{proposition:convolution}, we have
$$
\frac{a_1+a_2}{2} \exp(-L_{a_1,a_2})=\exp(-Y_{a_1,a_2}) \exp(-V_{a_1,a_2})\,.
$$
Then
$$
\left(\frac{a_1+a_2}{2}\right)^r {\left\{\exp(-L_{a_1,a_2})\right\}}^r={\left\{\exp(-Y_{a_1,a_2})\right\}}^r {\left\{\exp(-V_{a_1,a_2})\right\}}^r
$$
and
$$
\left(\frac{a_1+a_2}{2}\right)^r E({\left\{\exp(-L_{a_1,a_2})\right\}}^r)=E({\left\{\exp(-Y_{a_1,a_2})\right\}}^r) E({\left\{\exp(-V_{a_1,a_2})\right\}}^r)\,.
$$
The random variable $\exp(-L_{a_1,a_2})$ has a Log-logistic distribution and for any $r$ such that $|r|<1$,
$$
E({\left\{\exp(-L_{a_1,a_2})\right\}}^r)=\frac{r\pi}{\sin(r\pi)}\,.
$$
Moreover,
$$
E({\left\{\exp(-V_{a_1,a_2})\right\}}^r)=\int_0^1 \frac{a_1+a_2}{a_1-a_2} x^{r+\frac{a_1+a_2}{a_1-a_2}-1} dx=\frac{a_1+a_2}{(a_1-a_2)r+(a_1+a_2)}\,.
$$
Therefore,
$$
E({\left\{\exp(-Y_{a_1,a_2})\right\}}^r)=E({X_{a_1,a_2}}^r)=\left(\frac{a_1+a_2}{2}\right)^r \frac{r\pi}{\sin(r\pi)} \frac{(a_1-a_2)r+(a_1+a_2)}{a_1+a_2}\,.
$$

\vss We use the same type of reasoning for the case $a_1<a_2$.
\eproof

If $X_{a_1,a_2}^{b_1,b_2} \sim F_{a_1,a_2}^{b_1,b_2}$ then for any $r$ such that $|r|<b_2$
$$
E({X_{a_1,a_2}^{b_1,b_2}}^r)={b_1}^r \left(\frac{a_1+a_2}{2}\right)^\frac{r}{b_2} \frac{r\pi}{b_2\sin(\frac{r\pi}{b_2})} \left\{\frac{r(a_1-a_2)}{b_2(a_1+a_2)}+1\right\}\,.
$$

\subsection{Random sample generation}

When the distribution $F_0$ admits a density with respect to Lebesgue measure on $\mathbb{R}$, if we are able to generate a sample from $F_0$,
we can use the accept-reject algorithm to generate a sample from $F_{a_1,\ldots,a_q}$.

\newpage

Indeed, following the proof of Theorem \ref{theorem:expectation}, 
for any $(a_1,\ldots,a_q) \in {]0,+\infty[}^q$,
\begin{eqnarray*}
f_{a_1,\ldots,a_q}(x) \leq & \left\{\frac{\prod_{i=2}^q \max(1,a_i)}{{\min(1,\sum_{i=1}^q a_i/q)}^q}+ |\sum_{i=1}^q a_i-q| \frac{\prod_{i=2}^q \max(1,a_i)}{{\min(1,\sum_{i=1}^q a_i/q)}^{q+1}}\right. \\
                           & \left. +\frac{ \sum_{i=2}^q |1-a_i| \prod_{2 \leq j \leq q, j \neq i} \max(1,a_j)}{{\min(1,\sum_{i=1}^q a_i/q)}^q}\right\} f_0(x)\,.
\end{eqnarray*}

Moreover, if $a_1+\ldots+a_q \geq q$ and if for any integer $i$ such that $2 \leq i \leq q$, we have $a_i \leq 1$, then
$$
f_{a_1,\ldots,a_q}(x) \leq a_1 f_0(x)\,.
$$




\section{Conclusion}

We shown how to generalize the approach of \cite{Marshall:Olkin:1997} in order to obtain
more flexible families. We investigated some properties of the introduced distributions.
We are now working on the parameter estimation task. The preliminary results are
extremely encouraging, notably the application of our proposal on some real survival
datasets.

\newpage

\appendix

\section*{Appendix}

\subsection*{Proof of Theorem \ref{theorem:mainres}}

\vss The case $q=1$ is trivial, it corresponds to the Marshall and Olkin derivation.

\vss \textbf{Part one:}

\vss Let us first prove that for any $(a_1,a_2) \in {]0,+\infty[}^2$, the function $g_{a_1,a_2}$ from $[0,1]$ into $\mathbb{R}$ is increasing on $[0,1]$ and 
takes values in $[0,1]$.

\vss For any $u \in [0,1]$, we have,
$$
g_{a_1,a_2}'(u)=\frac{4 (a_1a_2u+a_1^2u-a_1a_2-a_1^2-2a_2u)}{(a_1u+a_2u-a_1-a_2-2u)^3}\,.
$$
Moreover,
$$
a_1a_2(u-1)+a_1^2(u-1)-2a_2u < 0 \,,
$$
$$
a_1(u-1)+a_2(u-1)-2u < 0 \,,
$$
and,
$$
{\left\{a_1(u-1)+a_2(u-1)-2u\right\}}^3 < 0\,.
$$
Therefore,
$$
g_{a_1,a_2}'(u) > 0\,.
$$
Thus, the function $g_{a_1,a_2}$ from $[0,1]$ to $\mathbb{R}$ is increasing on $[0,1]$.
As $g_{a_1,a_2}(0)=0$ and $g_{a_1,a_2}(1)=1$, the function $g_{a_1,a_2}$ from $[0,1]$ into $\mathbb{R}$ takes values in $[0,1]$.

\newpage

\vss \textbf{Part two:}

\vss Let us now prove by induction that for any integer $q$ such that $q\geq 2$ and any $(a_1,\ldots,a_q) \in {]0,+\infty[}^q$,
the function $g_{a_1,\ldots,a_q}$ from $[0,1]$ to $\mathbb{R}$ is increasing on $[0,1]$ and takes values in $[0,1]$.

\vss For any integer $q$ such that $q\geq 1$, we denote by $(H_q)$ the condition such that for any $(a_1,\ldots,a_q) \in {]0,+\infty[}^q$,
the function $g_{a_1,\ldots,a_q}$ from $[0,1]$ to $\mathbb{R}$ is increasing on $[0,1]$ and takes values in $[0,1]$.

\vss Let $q$ be an integer such that $q\geq 3$ and suppose that the condition $(H_{q-1})$ is verified. \\
In that case, for any $(u,a_1,\ldots,a_q) \in [0,1] \times {]0,+\infty[}^q$, we have,
$$
g_{a_1,\ldots,a_q}(u)=g_{\frac{a_1}{a_q},\ldots,\frac{a_{q-1}}{a_q},1}\left\{g_{a_q,\ldots,a_q}(u)\right\}
$$
and,
$$
g_{\frac{a_1}{a_q},\ldots,\frac{a_{q-1}}{a_q},1}(u)= \frac{q^q u \prod_{i=2}^{q-1} 
(\frac{a_i}{a_q}+u-\frac{a_i}{a_q} u)}{\left[\sum_{i=1}^{q-1} \frac{a_i}{a_q}+1-\left\{\sum_{i=1}^{q-1} \frac{a_i}{a_q}-(q-1)\right\}u\right]^q} \,,
$$
$$
g_{\frac{a_1}{a_q},\ldots,\frac{a_{q-1}}{a_q},1}(u)=\frac{q^q}{{(q-1)}^{q-1}} g_{\frac{a_1}{a_q},\ldots,\frac{a_{q-1}}{a_q}}(u) 
\frac{\left[\sum_{i=1}^{q-1} \frac{a_i}{a_q}-\left\{\sum_{i=1}^{q-1} \frac{a_i}{a_q}-(q-1)\right\}u\right]^{q-1}}
{\left[\sum_{i=1}^{q-1} \frac{a_i}{a_q}+1-\left\{\sum_{i=1}^{q-1} \frac{a_i}{a_q}-(q-1)\right\}u\right]^q} \,.
$$
Also,
$$
{\left[\sum_{i=1}^{q-1} \frac{a_i}{a_q}+1-\left\{\sum_{i=1}^{q-1} \frac{a_i}{a_q}-(q-1)\right\}u\right]}^{q}=\sum_{j=0}^{q} 
{q \choose j} {\left[\sum_{i=1}^{q-1} \frac{a_i}{a_q}-\left\{\sum_{i=1}^{q-1} \frac{a_i}{a_q}-(q-1)\right\}u\right]}^{q-j} \,,
$$
$$
\frac{\left[\sum_{i=1}^{q-1} \frac{a_i}{a_q}+1-\left\{\sum_{i=1}^{q-1} \frac{a_i}{a_q}-(q-1)\right\}u\right]^{q}}
{\left[\sum_{i=1}^{q-1} \frac{a_i}{a_q}-\left\{\sum_{i=1}^{q-1} \frac{a_i}{a_q}-(q-1)\right\}u\right]^{q-1}}=\sum_{j=0}^{q} 
{q \choose j}\left[\sum_{i=1}^{q-1} \frac{a_i}{a_q}-\left\{\sum_{i=1}^{q-1} \frac{a_i}{a_q}-(q-1)\right\}u\right]^{1-j} \,.
$$
For any $u \in [0,1]$, let us define,
$$
\eta_{a_1,\ldots,a_q}(u)=\frac{\left[\sum_{i=1}^{q-1} \frac{a_i}{a_q}+1-\left\{\sum_{i=1}^{q-1} \frac{a_i}{a_q}-(q-1)\right\}u\right]^{q}}
{\left[\sum_{i=1}^{q-1} \frac{a_i}{a_q}-\left\{\sum_{i=1}^{q-1} \frac{a_i}{a_q}-(q-1)\right\}u\right]^{q-1}}\,.
$$
We have,
$$
\eta_{a_1,\ldots,a_q}'(u)
$$
$$
=-\left\{\sum_{i=1}^{q-1} \frac{a_i}{a_q}-(q-1)\right\}-\sum_{j=2}^{q} {q \choose j} (1-j) \left\{\sum_{i=1}^{q-1} \frac{a_i}{a_q}-(q-1)\right\} 
{\left[\sum_{i=1}^{q-1} \frac{a_i}{a_q}-\left\{\sum_{i=1}^{q-1} \frac{a_i}{a_q}-(q-1)\right\}u\right]}^{-j}
$$
$$
=-\left\{\sum_{i=1}^{q-1} \frac{a_i}{a_q}-(q-1)\right\}\left\{1+\sum_{j=2}^{q} {q \choose j} (1-j) {\left[\sum_{i=1}^{q-1}
\frac{a_i}{a_q}-\left\{\sum_{i=1}^{q-1} \frac{a_i}{a_q}-(q-1)\right\}u\right]}^{-j}\right\}\,,
$$
and then,
$$
\eta_{a_1,\ldots,a_q}'(1)=-\left\{\sum_{i=1}^{q-1} \frac{a_i}{a_q}-(q-1)\right\}\left\{1+\sum_{j=2}^{q} {q \choose j} (1-j) {(q-1)}^{-j}\right\}\,.
$$
For any $x\in\mathbb{R}-\{0\}$, let us define, 
$$
r_q(x)=\sum_{j=2}^{q} {q \choose j} {x}^{-j+1}
$$
We have
$$
r_q(x)=x\left\{{\left(\frac{1}{x}+1\right)}^{q}-1-\frac{q}{x}\right\} \,,
$$
$$
r_q(x)=x{\left(\frac{1}{x}+1\right)}^{q}-x-q \,,
$$
and,
$$
r_q'(x)=\sum_{j=2}^{q} {q \choose j} \left(1-j\right) {x}^{-j}={\left(\frac{1}{x}+1\right)}^{q}-\frac{q}{x}
{\left(\frac{1}{x}+1\right)}^{q-1}-1={\left(\frac{1}{x}+1\right)}^{q-1}\left(\frac{1-q}{x}+1\right)-1 \,,
$$
$$
r_q'(q-1)+1=0 \,.
$$
Therefore,
$$
\eta_{a_1,\ldots,a_q}'(1)=0\,.
$$
Moreover, we have,
$$
\eta_{a_1,\ldots,a_q}''(u)={\left\{\sum_{i=1}^{q-1} \frac{a_i}{a_q}-(q-1)\right\}}^2 \sum_{j=2}^{q}
{q \choose j} (j-1) j \left[\sum_{i=1}^{q-1} \frac{a_i}{a_q}-\left\{\sum_{i=1}^{q-1} \frac{a_i}{a_q}-(q-1)\right\}u\right]^{-j-1}\,.
$$
Then clearly, $\eta_{a_1,\ldots,a_q}''(u) \geq 0$ and $\eta_{a_1,\ldots,a_q}'(u) \leq \eta_{a_1,\ldots,a_q}'(1)=0$.
Therefore, the function $u \mapsto \frac{1}{\eta_{a_1,\ldots,a_q}(u)}$ from $[0,1]$ to $\mathbb{R}_+$ is increasing on $[0,1]$. 
\\Using condition $(H_{q-1})$, the function $u \mapsto g_{\frac{a_1}{a_q},\ldots,\frac{a_{q-1}}{a_q}}(u)$ from $[0,1]$ to $\mathbb{R}$ is increasing
and positive. It follows that the function $g_{a_1,\ldots,a_q}$ from $[0,1]$ to $\mathbb{R}_+$  is increasing on $[0,1]$. 
\\Moreover, $g_{a_1,\ldots,a_q}(0)=0$ and $g_{a_1,\ldots,a_q}(1)=1$.
Thus, the condition $(H_q)$ is verified. \\ 
Therefore, we get the result by induction.

\vss \textbf{Part three:}

\vss Let us now prove that the function $F_{a_1,\ldots,a_q}$ from $\mathbb{R}$ to $[0,1]$ defines a probability distribution. \\
The function $F_{a_1,\ldots,a_q}$ from $\mathbb{R}$ to $\mathbb{R}$ takes its values in $[0,1]$, and it is a right continuous increasing function.
Moreover, $\lim_{x\rightarrow -\infty}F_{a_1,\ldots,a_q}(x)=0$ and $\lim_{x\rightarrow +\infty}F_{a_1,\ldots,a_q}(x)=1$.
\\Thus, it is a probability distribution.

\subsection*{Proof of Lemma \ref{lemma:dev1}}

Let $(a_1, \ldots,a_q) \in {]0,+\infty[}^q$ such that $a_1+\cdots+a_q > \frac{q}{2}$. For any $u\in[-1,1]$, we have,
$$
g_{a_1,\ldots,a_q}(u)=q^q \frac{u \prod_{i=2}^q \left\{a_i+(1-a_i)u\right\}}{{\left\{\sum_{i=1}^q a_i-(\sum_{i=1}^q a_i-q)u\right\}}^q} \,,
$$
$$
\frac{1}{\left\{\sum_{i=1}^{q} a_i-(\sum_{i=1}^{q} a_i-q)u\right\}^q}=\frac{1}{(\sum_{i=1}^q a_i)^q} \times \frac{1}{\left\{1-\frac{(\sum_{i=1}^q a_i-q)u}{\sum_{i=1}^q a_i}\right\}^q} \,,
$$
$$
-1 <  \frac{(\sum_{i=1}^q a_i-q)u}{\sum_{i=1}^q a_i} <1 \,.
$$
Then,
$$
\frac{1}{\left\{1-\frac{(\sum_{i=1}^q a_i-q)u}{\sum_{i=1}^q a_i}\right\}^q}=\frac{1}{(q-1)!} 
\sum_{k=0}^{\infty} (k+q-1) \cdots (k+1) {\left\{\frac{\left(\sum_{i=1}^q a_i-q\right)u}{\sum_{i=1}^q a_i}\right\}}^k\,.
$$
Moreover,
$$
u \prod_{i=2}^q \left\{a_i+(1-a_i)u\right\}= (\prod_{i=2}^q a_i) u \prod_{i=2}^q (1+\frac{1-a_i}{a_i}u)=(\prod_{i=2}^q a_i) u \prod_{i=1}^{q-1} (1+\frac{1-a_{i+1}}{a_{i+1}}u)\,,
$$
$$
u \prod_{i=2}^q \left\{a_i+(1-a_i)u\right\}=(\prod_{i=2}^q a_i) u \sum_{i=0}^{q-1} 
\sigma_i u^i=(\prod_{i=2}^q a_i) \sum_{i=0}^{q-1} \sigma_i u^{i+1}=(\prod_{i=2}^q a_i) \sum_{i=1}^q \sigma_{i-1} u^{i}=(\prod_{i=2}^q a_i) \sum_{j=1}^{\infty} \sigma_{j-1} u^{j}\,.
$$
Therefore,
$$
g_{a_1,\ldots,a_q}(u)=\frac{q^q}{(q-1)!} \frac{\prod_{i=2}^q a_i}{{(\sum_{i=1}^q a_i)}^q} (\sum_{j=1}^{\infty} \sigma_{j-1} u^{j}) \sum_{k=0}^{\infty} (k+q-1) \cdots (k+1) 
{\left\{\frac{\left(\sum_{i=1}^q a_i-q\right)u}{\sum_{i=1}^q a_i}\right\}}^k \,,
$$
$$
g_{a_1,\ldots,a_q}(u)=\frac{q^q}{(q-1)!} \frac{\prod_{i=2}^q a_i}{\left(\sum_{i=1}^q a_i\right)^q} \sum_{m=1}^{\infty} \left\{ \sum_{j+k=m, 1 \leq j \leq q} \sigma_{j-1} (k+q-1) \cdots (k+1) 
\left(\frac{\sum_{i=1}^q a_i-q}{\sum_{i=1}^q a_i}\right)^k \right\} u^m \,.
$$

\subsection*{Proof of Lemma \ref{lemma:dev2}}

Let $(a_1, \ldots,a_q) \in {]0,+\infty[}^q$ such that $a_1+\cdots+a_q < 2q$ and $u\in[0,2]$. We define $w=u-1$, we get that $w \in [-1,1]$ and,
$$
g_{a_1,\ldots,a_q}(u)=q^q \frac{u \prod_{i=2}^q (a_i+u-a_iu)}{{\left\{\sum_{i=1}^q a_i-(\sum_{i=1}^q a_i-q)u\right\}}^q} \,,
$$
$$
g_{a_1,\ldots,a_q}(u)=q^q \frac{(1+w) \prod_{i=2}^q \left\{a_i+1+w-a_i(1+w)\right\}}{{\left\{\sum_{i=1}^q a_i-(\sum_{i=1}^q a_i-q)(1+w)\right\}}^q} \,,
$$
$$
g_{a_1,\ldots,a_q}(u)=q^q \frac{(1+w) \prod_{i=2}^q \left\{1+(1-a_i)w\right\}}{{\left\{q-(\sum_{i=1}^q a_i-q)w\right\}}^q} \,,
$$
$$
g_{a_1,\ldots,a_q}(u)=\frac{(1+w) \prod_{i=2}^q \left\{1+(1-a_i)w\right\}}{{(1-\frac{\sum_{i=1}^q a_i-q}{q}w)}^q} \,.
$$
Moreover, $-1<\frac{\sum_{i=1}^q a_i-q}{q} <1$, then $-1<\frac{\sum_{i=1}^q a_i-q}{q} w <1$ and,
$$
\frac{1}{{(1-\frac{\sum_{i=1}^q a_i-q}{q}w)}^q}=\frac{1}{(q-1)!} \sum_{k=0}^{\infty} (k+q-1) \cdots (k+1) {\left\{\frac{(\sum_{i=1}^q a_i-q)w}{q}\right\}}^k\,.
$$
Also,
$$
\prod_{i=2}^q \left\{1+(1-a_i)w\right\}=\sum_{m=0}^{q-1} \sigma_m w^m=\sum_{m=0}^{\infty} \sigma_m w^m \,,
$$
$$
(1+w) \prod_{i=2}^q \left\{1+(1-a_i)w\right\}=\sum_{m=0}^{\infty} \sigma_m w^m+\sum_{m=0}^{\infty} \sigma_m w^{m+1} \,,
$$
$$
(1+w) \prod_{i=2}^q \left\{1+(1-a_i)w\right\}=1+\sum_{m=1}^{\infty} (\sigma_m+\sigma_{m-1}) w^m \,.
$$
Then,
$$
g_{a_1,\ldots,a_q}(u)=\frac{1}{(q-1)!} \left\{1+\sum_{m=1}^{\infty} (\sigma_m+\sigma_{m-1}) w^m\right\}
\left[\sum_{k=0}^{\infty} (k+q-1) \cdots (k+1) \left\{\frac{(\sum_{i=1}^q a_i-q)w}{q}\right\}^k\right] \,,
$$
\begin{eqnarray*}
g_{a_1,\ldots,a_q}(u)= & \frac{1}{(q-1)!} \left\{\sum_{k=0}^{\infty} (k+q-1) \cdots (k+1) {(\frac{\sum_{i=1}^q a_i-q}{q})}^k w^k \right. \\
                       & \left. +\sum_{m=1}^{\infty} \sum_{k+j=m,1 \leq j \leq q} (k+q-1) \ldots (k+1) {(\frac{\sum_{i=1}^q a_i-q}{q})}^k (\sigma_j+\sigma_{j-1}) w^m\right\} \,,
\end{eqnarray*}
\begin{eqnarray*}
g_{a_1,\ldots,a_q}(u)= & 1+\frac{1}{(q-1)!} \left[\sum_{m=1}^{\infty} \left\{(m+q-1) \ldots (m+1) {(\frac{\sum_{i=1}^q a_i-q}{q})}^m \right.\right. \\
                       & \left.\left. +\sum_{k+j=m,1 \leq j \leq q} (k+q-1) \ldots (k+1) {(\frac{\sum_{i=1}^q a_i-q}{q})}^k (\sigma_j+\sigma_{j-1})\right\} w^m\right] \,.
\end{eqnarray*}

\subsection*{Proof of Theorem \ref{theorem:expectation}}

\vss \textbf{Part one:}

\vss Let us prove first that for any integer $q$ such that $q \geq 1$, any $(a_1,\ldots,a_q) \in {]0,+\infty[}^q$ and any $u \in [0,1]$, we have,
$$
g_{a_1,\ldots,a_q}'(1)=a_1 \,,
$$
$$
g_{a_1,\ldots,a_q}'(u) \geq 0 \,,
$$
$$
g_{a_1,\ldots,a_q}'(u) 
$$
$$
\leq  \frac{\prod_{i=2}^q \max(1,a_i)}{{\min(1,\frac{\sum_{i=1}^q a_i}{q})}^q}
+ |\sum_{i=1}^q a_i-q| \frac{ \prod_{i=2}^q \max(1,a_i)}{{\min(1,\frac{\sum_{i=1}^q a_i}{q})}^{q+1}}
+\frac{ \sum_{i=2}^q |1-a_i| \prod_{2 \leq j \leq q, j \neq i} \max(1,a_j)}{{\min(1,\frac{\sum_{i=1}^q a_i}{q})}^q} \,.
$$

\vss For any $u \in [0,1]$, we have,
\begin{eqnarray*}
g_{a_1,\ldots,a_q}'(u)= & q^q \frac{\prod_{i=2}^q (a_i+u-a_i u)}{{\left\{\sum_{i=1}^q a_i-(\sum_{i=1}^q a_i-q)u\right\}}^q} \\
& + q^{q+1} (\sum_{i=1}^q a_i-q)\frac{u \prod_{i=2}^q (a_i+u-a_i u)}{{\left\{\sum_{i=1}^q a_i-(\sum_{i=1}^q a_i-q)u\right\}}^{q+1}}\\
& +q^q \frac{u \sum_{i=2}^q (1-a_i) \prod_{2 \leq j \leq q, j \neq i} (a_j+u-a_j u)}{{\left\{\sum_{i=1}^q a_i-(\sum_{i=1}^q a_i-q)u\right\}}^q}\,.
\end{eqnarray*}
Then,
$$
g_{a_1,\ldots,a_q}'(1)=1+\sum_{i=1}^q a_i -q+\sum_{i=2}^q(1-a_i)=a_1\,.
$$
For any $u \in [0,1]$, we have
\begin{eqnarray*}
g_{a_1,\ldots,a_q}'(u) \leq & |q^q \frac{\prod_{i=2}^q (a_i+u-a_i u)}{{\left\{ \sum_{i=1}^q a_i-(\sum_{i=1}^q a_i-q)u \right\}}^q}| \\
& + |q^{q+1} (\sum_{i=1}^q a_i-q)\frac{u \prod_{i=2}^q (a_i+u-a_i u)}{{\left\{\sum_{i=1}^q a_i-(\sum_{i=1}^q a_i-q)u\right\}}^{q+1}}|\\
& +|q^q \frac{u \sum_{i=2}^q (1-a_i) \prod_{2 \leq j \leq q, j \neq i} (a_j+u-a_j u)}{{\left\{\sum_{i=1}^q a_i-(\sum_{i=1}^q a_i-q)u\right\}}^q}|\,.
\end{eqnarray*}
Moreover, for any integer $2 \leq i \leq q$
$$
0 \leq \min(1,a_i) \leq a_i+u-a_i u \leq \max(1,a_i)\,,
$$
$$
0< \min(q,\sum_{i=1}^q a_i) \leq \sum_{i=1}^q a_i-(\sum_{i=1}^q a_i-q)u \leq \max(q,\sum_{i=1}^q a_i)\,.
$$
Therefore,
\begin{eqnarray*}
g_{a_1,\ldots,a_q}'(u) \leq & q^q \frac{\prod_{i=2}^q \max(1,a_i)}{{\min(q,\sum_{i=1}^q a_i)}^q} \\
& + q^{q+1} |\sum_{i=1}^q a_i-q| \frac{ \prod_{i=2}^q \max(1,a_i)}{{\min(q,\sum_{i=1}^q a_i)}^{q+1}}\\
& +q^q \frac{ \sum_{i=2}^q |1-a_i| \prod_{2 \leq j \leq q, j \neq i} \max(1,a_j)}{{\min(q,\sum_{i=1}^q a_i)}^q}\,,
\end{eqnarray*}
$$
g_{a_1,\ldots,a_q}'(u) \leq \frac{\prod_{i=2}^q \max(1,a_i)}{{\min(1,\frac{\sum_{i=1}^q a_i}{q})}^q}
+ |\sum_{i=1}^q a_i-q| \frac{ \prod_{i=2}^q \max(1,a_i)}{{\min(1,\frac{\sum_{i=1}^q a_i}{q})}^{q+1}}
+\frac{ \sum_{i=2}^q |1-a_i| \prod_{2 \leq j \leq q, j \neq i} \max(1,a_j)}{{\min(1,\frac{\sum_{i=1}^q a_i}{q})}^q}\,.
$$
As the function $g_{a_1,\ldots,a_q}$ from $[0,1]$ to $\mathbb{R}$ is increasing and differentiable on $[0,1]$, for any $u \in [0,1]$,
$$
g_{a_1,\ldots,a_q}'(u) \geq 0\,.
$$

\vss \textbf{Part two:}

\vss Let us suppose that $\int_{\mathbb{R}} |w(x)| f_0(x) dx \in \mathbb{R}_+$. We have
$$
E(|w \circ X_{a_1,\ldots,a_q}|)=\int_{\mathbb{R}} |w(x)| g_{a_1,\ldots,a_q}'\left\{F_0(x)\right\} f_0(x) dx\,.
$$
Using the result given in Part one, for any $u\in[0,1]$,
\begin{eqnarray*}
E(|w \circ X_{a_1,\ldots,a_q}|) \leq & \left\{\frac{\prod_{i=2}^q \max(1,a_i)}{{\min(1,\frac{\sum_{i=1}^q a_i}{q})}^q} \right.\\
                                 & +|\sum_{i=1}^q a_i-q| \frac{ \prod_{i=2}^n \max(1,a_i)}{{\min(1,\frac{\sum_{i=1}^q a_i}{q})}^{q+1}} \\
                                 & \left. +\frac{ \sum_{i=2}^q |1-a_i| \prod_{2 \leq j \leq q, j \neq i} \max(1,a_j)}{{\min(1,\frac{\sum_{i=1}^q a_i}{q})}^q}\right\} E(|w \circ X_0|)\,.
\end{eqnarray*}
Therefore,
$$
E(|w \circ X_{a_1,\ldots,a_q}|) \in \mathbb{R}_+\,.
$$
By the corollary~\ref{corollary:signe_coefficients}, if $q\geq 1$ and $a_1,\ldots,a_q$ be strictly positive real numbers such that
$a_1+\cdots+a_q \geq q$ and $a_i\leq 1$ for any $2 \leq i \leq q$, then we have that for any integer $m$ which is greater than or equal to one and any $u\in[0,1]$,
$$g_{a_1,\ldots,a_q}''(u) \geq 0$$ 
Therefore,
$$
g_{a_1,\ldots,a_q}'(u) \leq g_{a_1,\ldots,a_q}'(1)\,,
$$
$$
E(|w \circ X_{a_1,\ldots,a_q}|) \leq a_1 E(|w \circ X_0|)\,.
$$

\end{document}